\documentclass{czjphys}         

\usepackage{graphicx}
\usepackage{amssymb,amsmath,times}
\usepackage{epsfig}
\usepackage{units}
\usepackage{floatflt}
\usepackage{wrapfig}
\begin{document}
\title{Muon production in extensive air showers and its relation to hadronic interactions}  
%
\authori{C. Meurer, J. Bl\"umer, R. Engel, A. Haungs, M. Roth }      
\addressi{Forschungszentrum Karlsruhe GmbH, Postfach 3640, D-76021 Karlsruhe, Germany}
\authorii{}     \addressii{}
\authoriii{}    \addressiii{}
\authoriv{}     \addressiv{}
\authorv{}      \addressv{}
\authorvi{}     \addressvi{}
%
\headauthor{C. Meurer et al.}            
\headtitle{Muon production in extensive air showers and its relation to hadronic interactions
}             
\lastevenhead{C. Meurer et al.: Muon production in extensive air showers and its relation to hadronic interactions
} 
\pacs{13.85.Tp, 96.40.Pq}
\keywords{cosmic rays, extensive air showers, muon and hadron production, fixed target experiments} 
\refnum{A}
\daterec{XXX}    
\issuenumber{0}  \year{2001}
\setcounter{page}{1}
\maketitle

\begin{abstract}
In this work, the relation between muon production in extensive air showers
and features of hadronic multiparticle production at low energies is studied.
Using CORSIKA, we determine typical energies and phase space regions of secondary
particles which are important for muon production in extensive air
showers and confront the results with existing fixed target measurements. Furthermore 
possibilities to measure relevant quantities of hadron production in existing and 
planned accelerator experiments are discussed.
\end{abstract}

\section{Introduction}
The energy spectrum and composition of the cosmic rays with energies above $10^{15}$\,eV
are typically derived from measurements of the number of electrons and muons produced in
extensive air showers (EAS) at ground. 
However, the results of such a shower analysis are strongly dependent on
the hadronic interaction models used for simulating reference
showers \cite{kascade_holger}.  Therefore it is important to study in
detail the role of hadronic interactions and in particular the energy
and secondary particle phase space regions that are most important for
the observed characteristics of EAS.

The electromagnetic component of a shower is well determined by the
depth of maximum and the energy of the shower. Due to the electromagnetic cascade, 
having a short radiation length of \unit[$\sim 36$]{g/cm$^2$}, any information on 
the initial distribution of photons produced in $\pi^0$ decays is lost. 
Therefore the electromagnetic shower component depends on the primary
particle type only through the depth of shower maximum.
In contrast, the muon component is very sensitive to
the characteristics of hadronic interactions.  Once the hadronic shower
particles have reached an energy at which charged pions and kaons decay,
they produce muons which decouple from the shower cascade. The muons
propagate to the detector with small energy loss and deflection and
hence carry information on hadronic interactions in EAS.
Due to the competition between interaction and decay, most of the muons
are decay products of mesons that are produced in low-energy
interactions. Therefore it is not surprising that muons in EAS are
particularly sensitive to hadronic multiparticle production at low energy
\cite{EngelISMD1999, Drescher2003}. Recent model studies show that even at
ultra-high shower energies the predictions on the lateral distribution
of shower particles depend strongly on the applied low-energy interaction model
\cite{Drescher2004, Heck2003}.


\section{Muon production in extensive air showers}

Motivated by the measurement conditions of the KASCADE array \cite{kascadeNIM}, we
consider showers with a primary energy of $10^{15}$~eV and apply a muon detection threshold of
\unit[250]{MeV}.  Using a modified version of the simulation package
CORSIKA \cite{CORSIKA} we have simulated two samples of 1500 vertical
and inclined ($60^{\circ}$) proton and 500 iron induced showers.  Below
\unit[80]{GeV} the low-energy hadronic interaction model GHEISHA 2002
\cite{GHEISHA} and above \unit[80]{GeV} the high-energy model QGSJET 01
\cite{QGSJET} are applied. In the following vertical as well as $60^{\circ}$ inclined 
proton showers are discussed. The results are very similar for iron induced showers.   

\begin{figure}[h]
\begin{minipage}[t]{0.49\textwidth}
        \centering
        \includegraphics[width=\textwidth,bb=  10 20 511 350,clip]{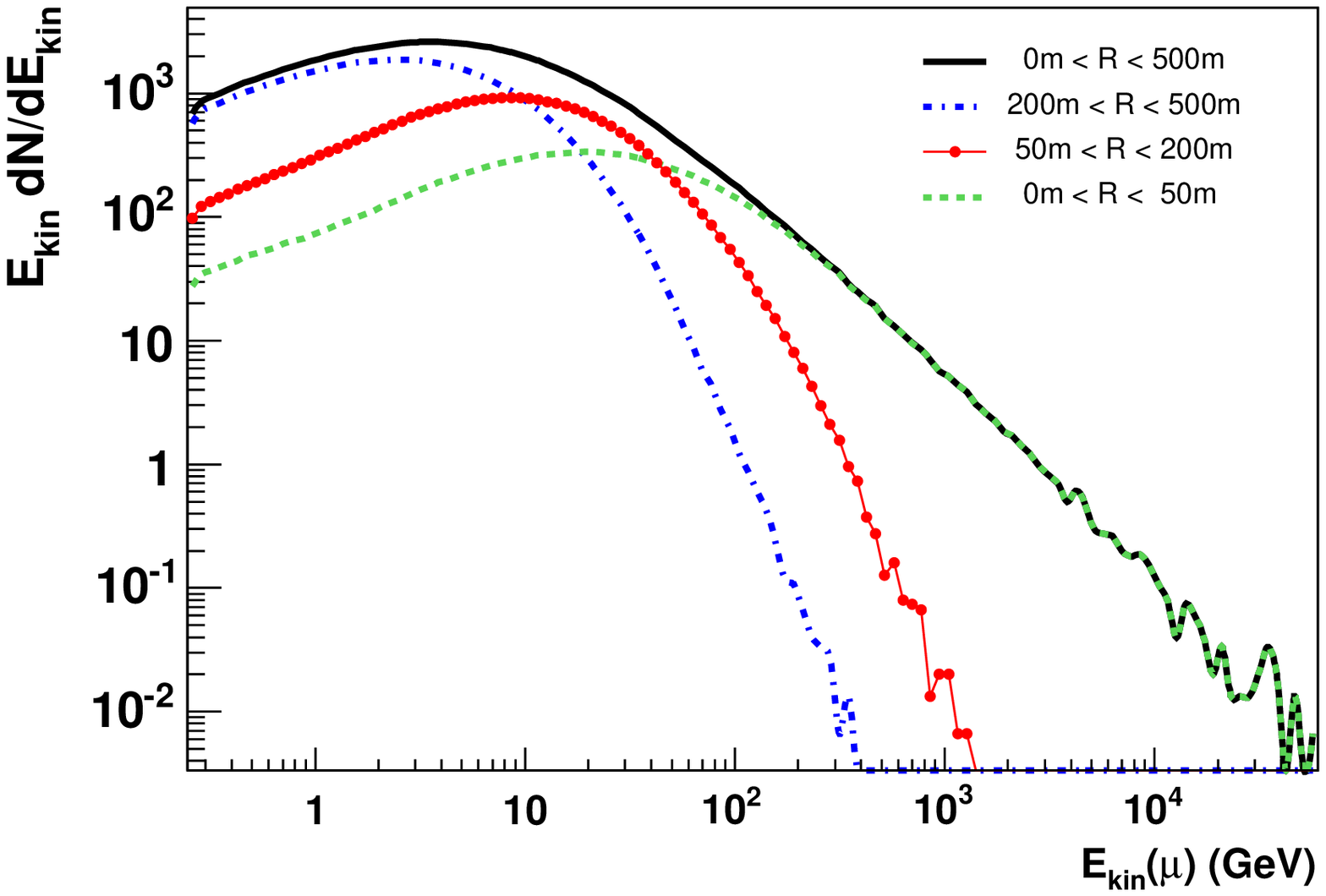}
        \caption{Simulated energy distribution of muons for different lateral distances in vertical proton induced showers with a primary energy of $10^{15}$eV.}
        \label{Edis_md}
\end{minipage}
\hfill
\begin{minipage}[t]{0.49\textwidth}
        \centering
        \includegraphics[width=\textwidth,bb=  10 20 511 350,clip]{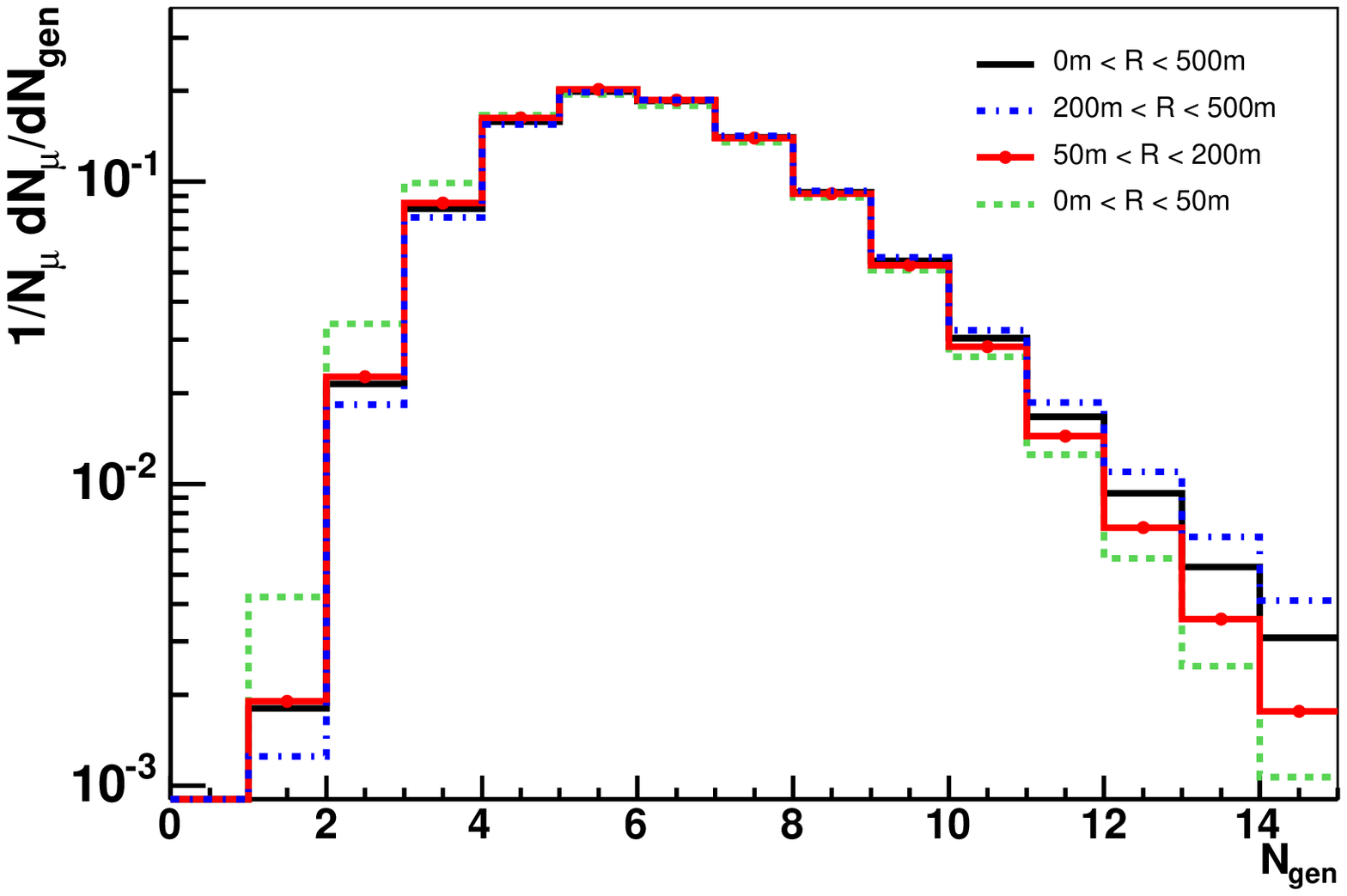}
        \caption{Number of generations before producing a muon visible at ground level (shown for various lateral distances).}
        \label{Ngen}
\end{minipage}
\end{figure}

In Fig.~\ref{Edis_md} the energy distribution of muons at detector level
(\unit[1030]{g/cm$^2$}) is shown for several lateral distance ranges. The
maximum of this distribution shifts to lower energies for larger lateral
distances.  Most likely five consecutive hadronic interactions
(number of generations) take place before a hadron decays into a muon,
see Fig.~\ref{Ngen}.  Here and in the following we consider only those
muons that reach the ground level with an energy above the detection
threshold.  The number of generations show only a weak dependence on
the lateral distance.  

To study the hadronic {\it ancestors} of muons in
EAS, we introduce the terms {\it grandmother} and {\it mother particle}
for each observed muon. The grandmother particle is the hadron inducing
the {\it last} hadronic interaction that finally leads to a meson
(mother particle) which decays into the corresponding muon.  Most of the
grandmother and mother particles are pions, but also about 20\% of the
grandmother particles are nucleons and a few are kaons. Details of the
composition of mother and grandmother particles are given in
Tab.~\ref{ID}.  

\begin{table}
\vspace*{-4mm}
\caption{\label{ID} Particle types of mother and grandmother particles 
in a vertical proton induced shower at $10^{15}$eV.}
\vspace*{2mm}
\centering
\begin{tabular}{|c|c|c|}
\hline          & mother  & grandmother \\
\hline
\hline pions    & 89.2\% &  72.3\%  \\
\hline kaons    & 10.5\% &   6.5\%  \\
\hline nucleons & -      &  20.9\%  \\
\hline
\end{tabular}
\end{table}


\section{Relevant energy range}

\begin{figure}[h!]
\begin{minipage}[t]{0.49\textwidth}
\centering
\includegraphics[width=\textwidth, bb=  10 20 511 345,clip]{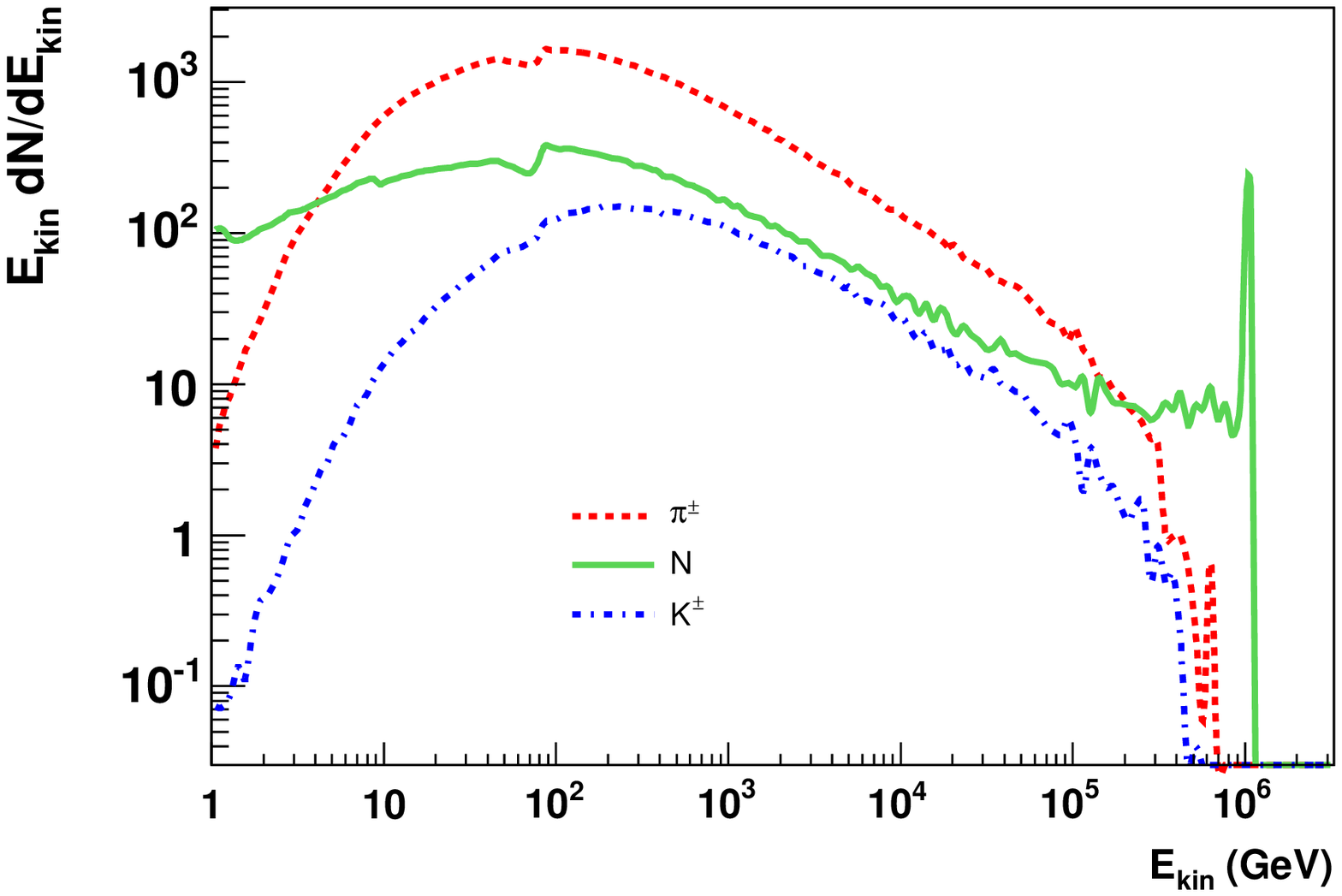}
\end{minipage}
\hfill
\begin{minipage}[t]{0.49\textwidth}
\centering
\includegraphics[width=\textwidth, bb=  10 20 511 345,clip]{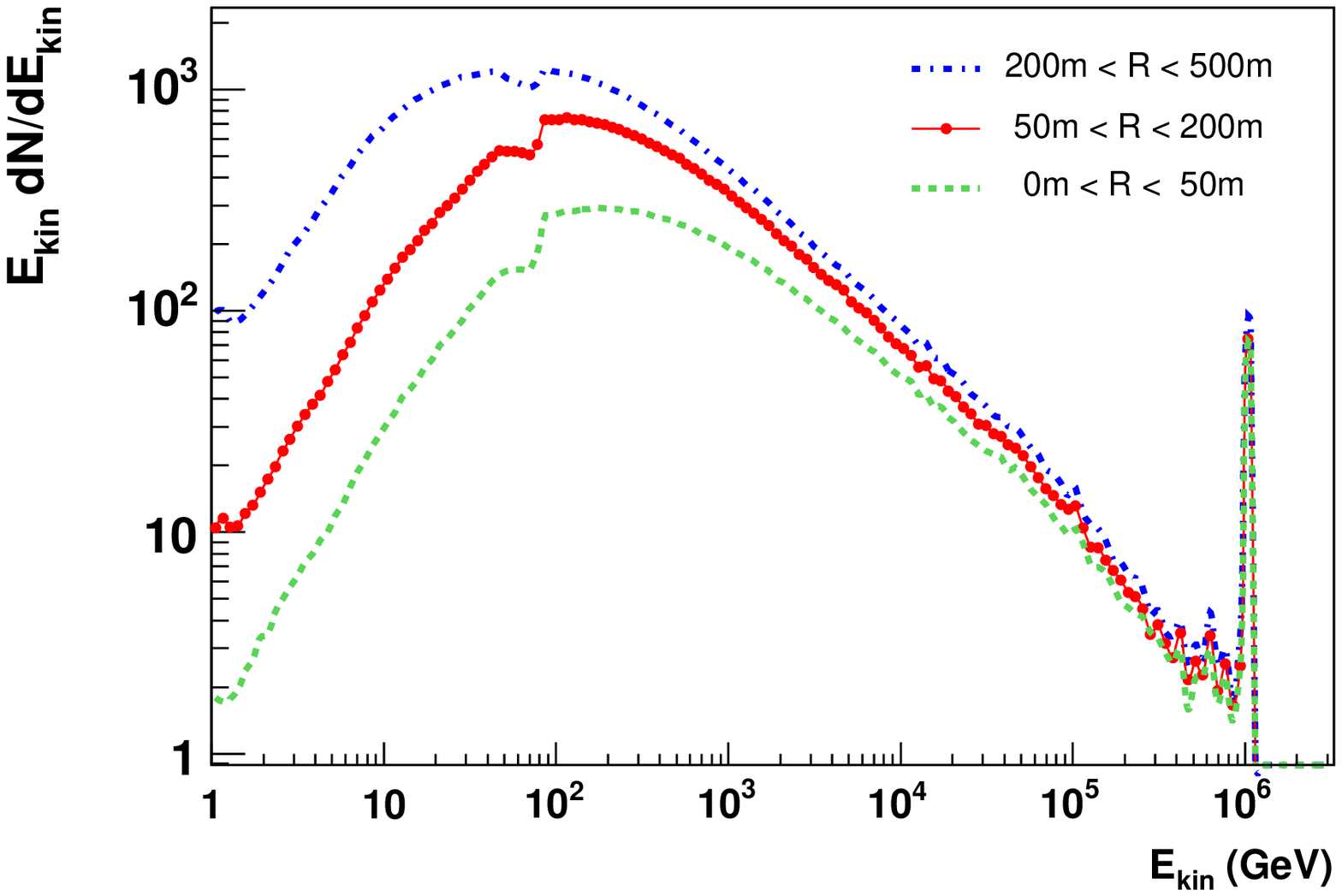} 
\end{minipage}
\caption{\label{Edis_gm} Energy distribution of grandmother particles in vertical proton showers. 
Left panel: different grandmother particle types for a muon lateral distance range of \unit[0-500]{m} at ground level. Right panel: different lateral distances, all particle types are summed up.}
\end{figure}
\begin{figure}[h!]
\begin{minipage}[t]{0.49\textwidth}
\includegraphics[width=\textwidth, bb=  10 20 511 345,clip]{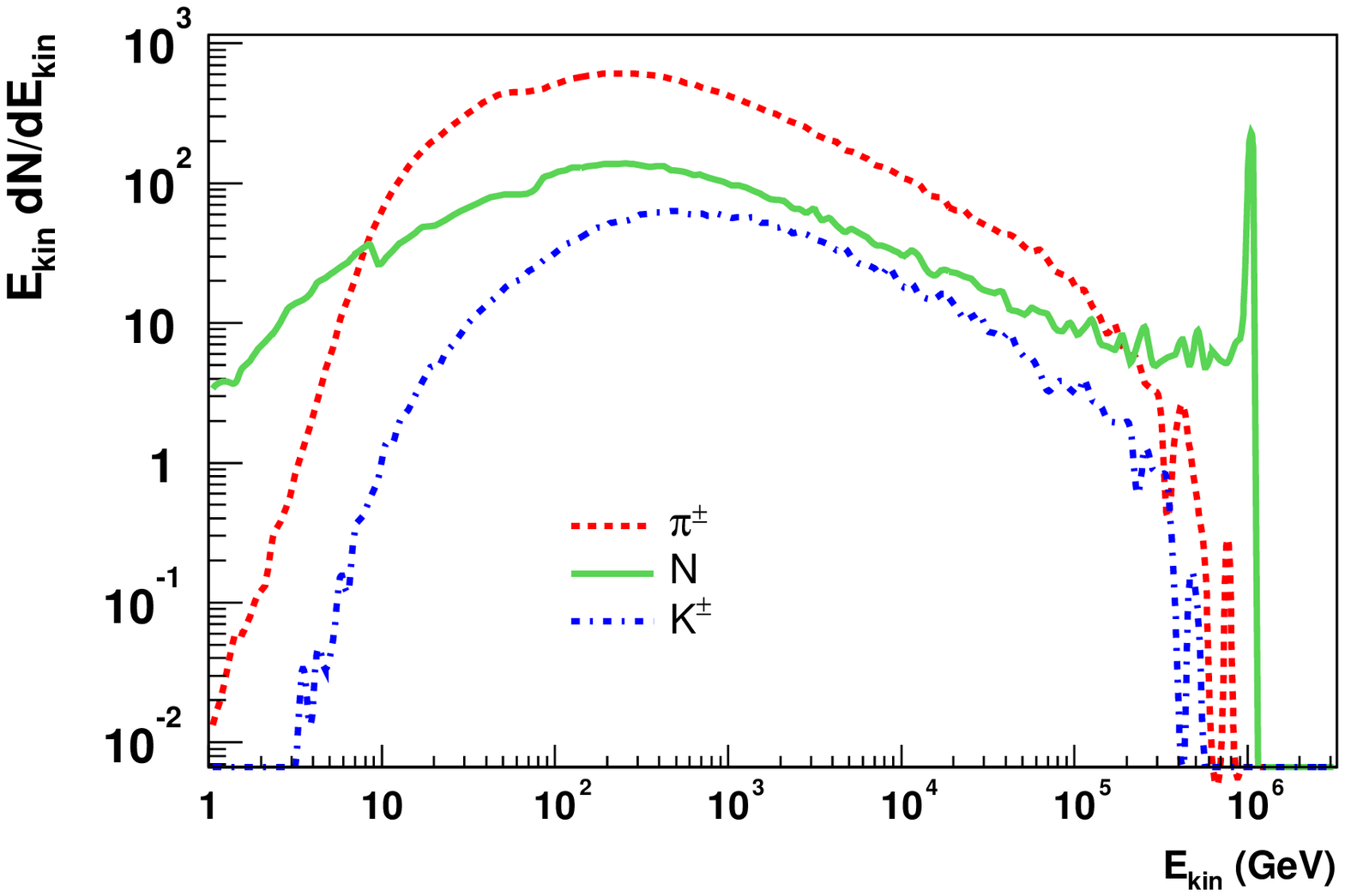}
\end{minipage}
\hfill
\begin{minipage}[t]{0.49\textwidth}
\includegraphics[width=\textwidth, bb=  10 20 511 345,clip]{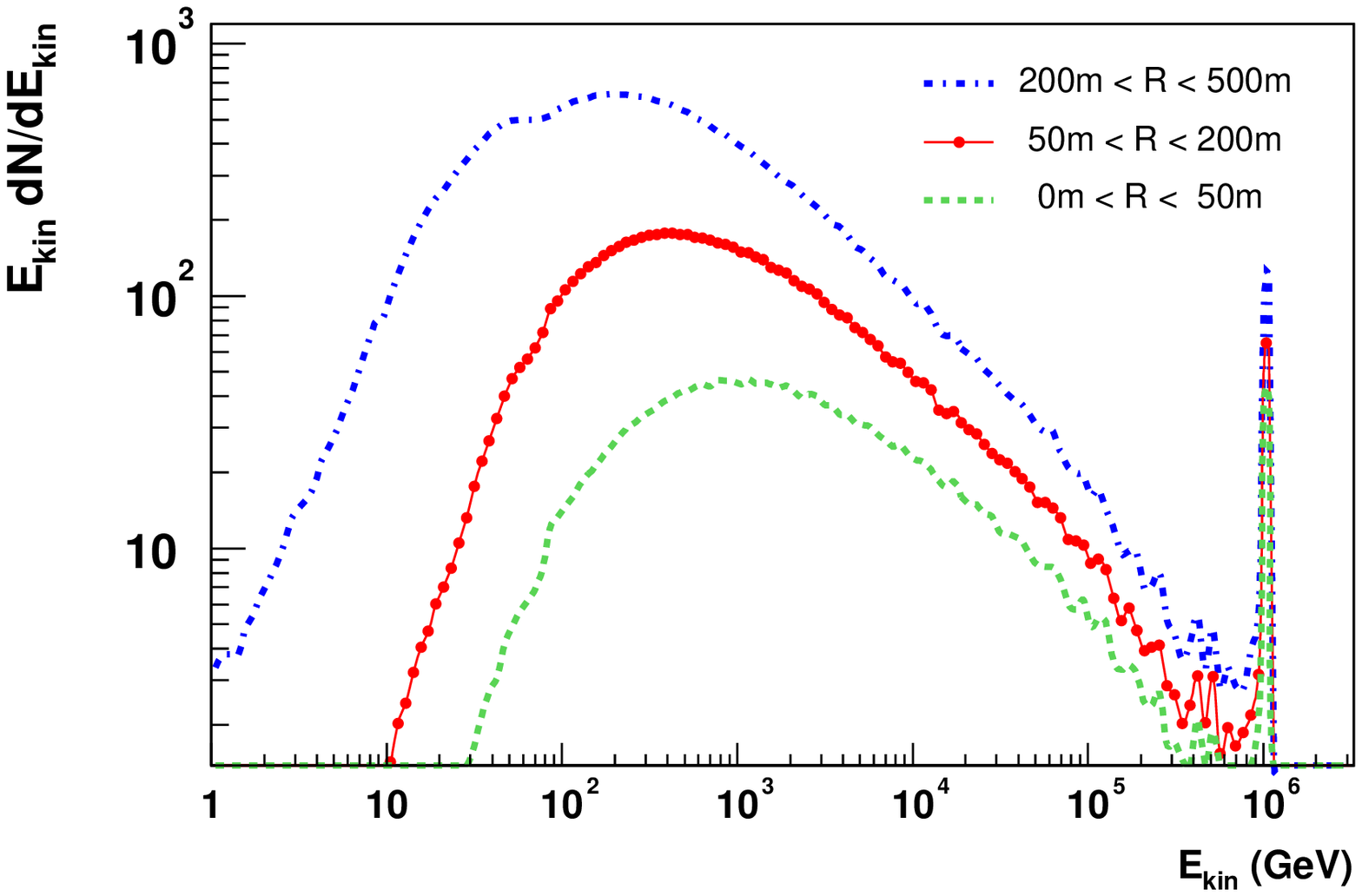} 
\end{minipage}
\caption{\label{p6015_Edis_gm} Same plots as in Fig.\ref{Edis_gm}, but for $60^{\circ}$ inclined proton showers. 
}
\end{figure}

The energy spectra of different grandmother particles produced in vertical proton showers 
are shown in Fig.~\ref{Edis_gm} (left). They cover a large energy range up to the
primary energy with a maximum at about \unit[100]{GeV}. The maximum shifts to higher energies 
(several \unit[100]{GeV}) for $60^{\circ}$ inclined proton showers, see Fig.~\ref{p6015_Edis_gm} (left).
The peak at \unit[$10^{6}$]{GeV} in the nucleon energy spectrum shows that also a
fraction of muons stems from decays of mesons produced in the first
interaction in a shower.  Furthermore, the step at \unit[80]{GeV}
clearly indicates a mismatch between the predictions of the low-energy
model GHEISHA and the high-energy model QGSJET.  In Fig.~\ref{Edis_gm}
(right) the grandmother particle energy spectrum is shown for different
ranges of lateral muon distance.  The maximum shifts with larger lateral
distance to lower energies. The same behaviour is visible for inclined showers, see 
Fig.~\ref{p6015_Edis_gm} (right).

Comparing the {\it last} interaction in EAS
with collisions studied at accelerators, one has to keep in mind that the
grandmother particle corresponds to the beam particle and the mother
particle is equivalent to a secondary particle produced in e.g. a
minimum bias p-N interaction. The most probable energy of the
grandmother particle is within the range of beam energies of fixed
target experiments e.g. at the PS and SPS accelerators at CERN.

\section{Relevant phase space regions}


The further study of the relevant phase space of the mother particles 
is done for two different grandmother energy ranges and 
muon lateral distance ranges at ground level, see Tab.~\ref{range}. The lateral
distance ranges are chosen to resemble typical lateral distances
measured at \mbox{KASCADE} and KASCADE-Grande, respectively \cite{KASCADE_C2CR05}. 
Motivated by the availability of protons as beam particles at accelerators we consider only those {\it last} interactions in EAS that are initiated by nucleons. 
Exemplary, the results of this analysis are only shown for vertical proton showers.

\begin{figure}[ht]
\begin{minipage}[ht]{0.49\textwidth}  
\centering
\includegraphics[width =\textwidth, bb=  10 20 511 345,clip]{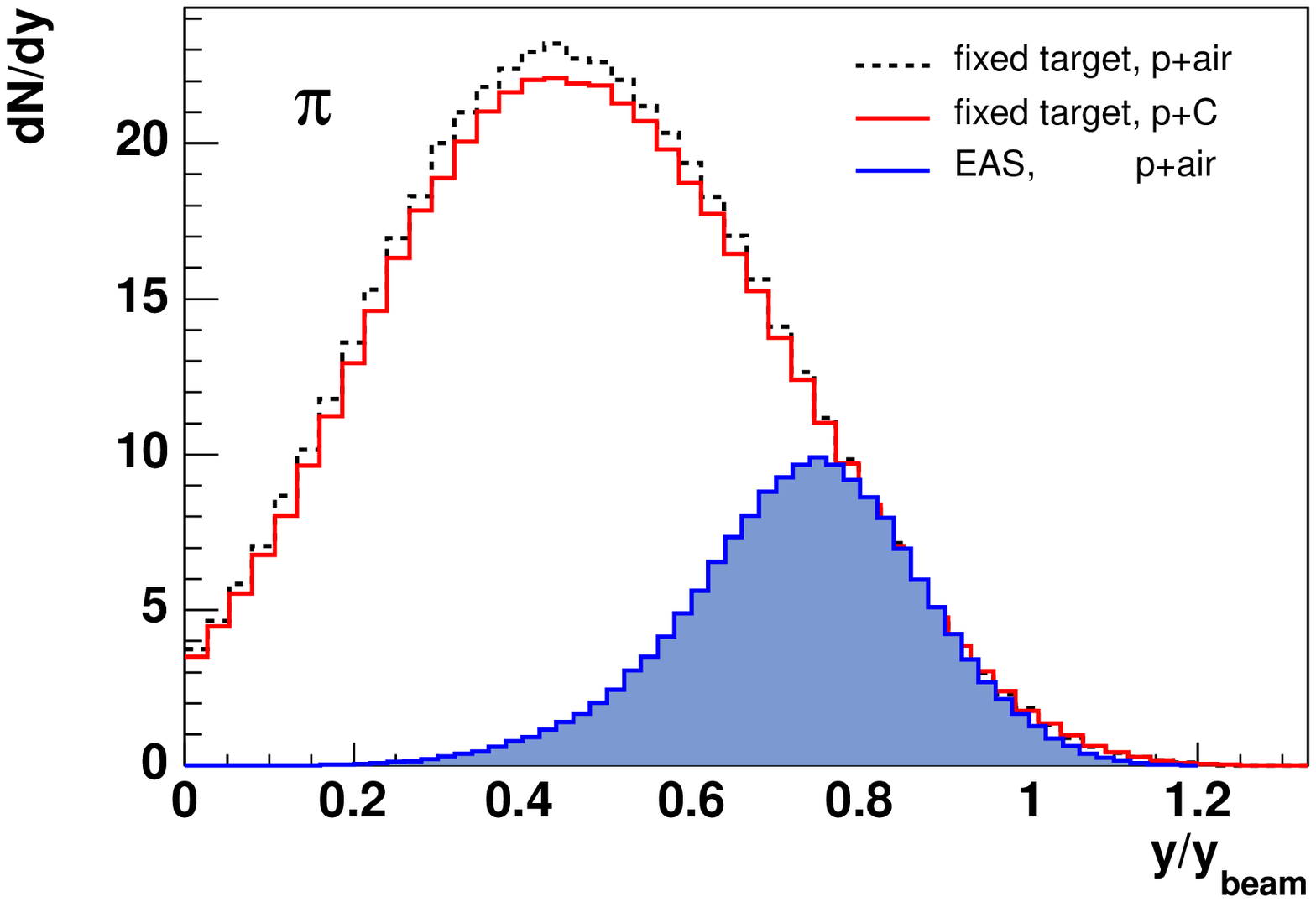}
\end{minipage}
\hfill
\begin{minipage}[ht]{0.49\textwidth}
\centering
\includegraphics[width =\textwidth, bb=  10 20 511 345,clip]{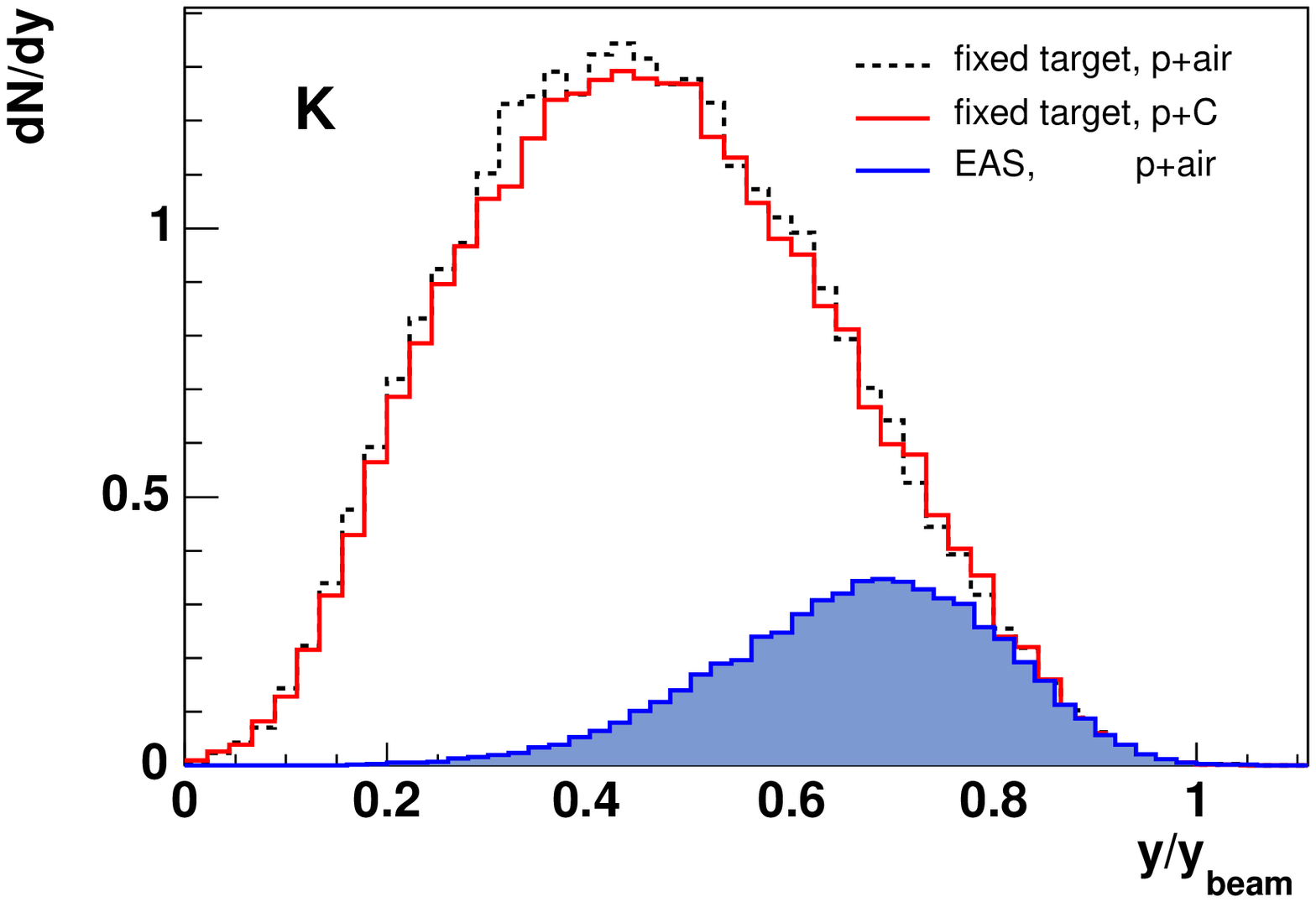}
\end{minipage}

\caption{\label{Rap} 
Rapidity distributions of mother particles (filled
curves) compared with rapidity distributions of secondary particles in
simulated single p+C (solid line) and simulated p+air (dashed line)
collisions. Left panel: pions, right panel: kaons.  The energy range of the
grandmother particle is limited to \unit[80-400]{GeV} and the lateral
distance of the muons to \unit[50-200]{m} to match experimentally
accessible regions. The fixed target collision simulation is done at
\unit[160]{GeV}, corresponding approximately to the mean grandmother
energy.  The rapidity is normalized to the rapidity of the beam and
grandmother particles, respectively.
}
\end{figure}

In Fig.~\ref{Rap} the rapidity spectra of mother particles (left: pions, right: kaons) are
compared to the spectra of secondary particles of minimum bias proton-carbon
 and proton-air collisions with a fixed energy simulated with QGSJET labeled as {\it fixed target}. 
The spectra of mother particles in air showers are scaled to fit the falling tail of
the fixed-energy collision spectra. No significant differences are found
comparing the rapidity distributions of secondary particles in
proton-carbon and proton-air collisions. 
As a consequence of the different selection criteria, the forward hemisphere in the 
mother rapidity spectra is clearly favoured compared to the spectra 
of secondaries in minimum bias collisions. The reason for this behaviour is the fact 
that the secondary particles (pions, kaons) are measured directly in fixed target experiments 
whereas in EAS an additional condition for the mother mesons is applied. 
In order to get the information of the mother meson, it has first to decay into a muon which is 
detectable. 
At low rapidity the mother mesons are missing in the EAS distribution 
because the energy of the daughter muon is lower than the applied detection threshold or 
the muon decays and it is not visible. 
The reason for the missing pions in the EAS distribution at the highest rapidities is 
the interplay between decay and further interactions, that means the pions do not decay but 
undergo further interactions. 
This effect is not much pronounced for the kaon distribution, because 
the energy, for which the probability of decay and interaction is of the same order, is eight times 
higher for kaons than for pions.

\vspace*{-2mm}
\begin{table}
\caption{\label{range} Energy and lateral distance ranges used for this analysis.
}
\vspace*{2mm}
\centering
\begin{tabular}{|c|c|c|}
\hline energy range             & average energy          & lateral distance range \\
\hline
\hline \unit[80-400]{GeV}       & \unit[160]{GeV}         &  \unit[50-200]{m}  \\
\hline \unit[30-60]{GeV}        & \unit[40]{GeV}          &  \unit[200-600]{m}  \\
\hline
\end{tabular}
\vspace{0.5 cm}
\end{table}

\clearpage

\begin{figure}[h+t!]
\begin{minipage}[t]{0.49\textwidth}
\centering
\includegraphics[width =\textwidth, bb=  10 20 519 345,clip]{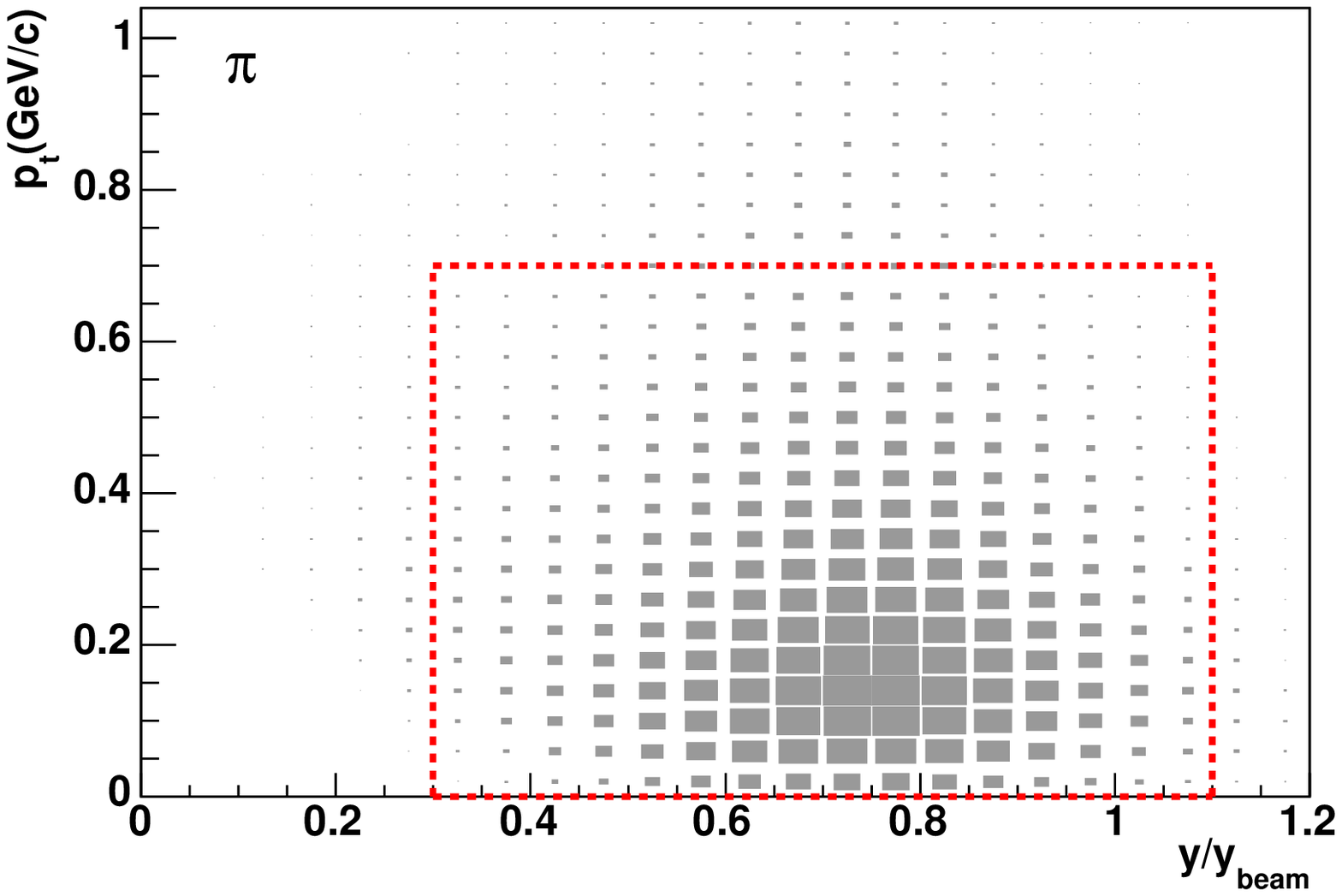}
\end{minipage}
\hfill
\begin{minipage}[t]{0.49\textwidth}
\centering
\includegraphics[width =\textwidth, bb=  10 20 519 345,clip]{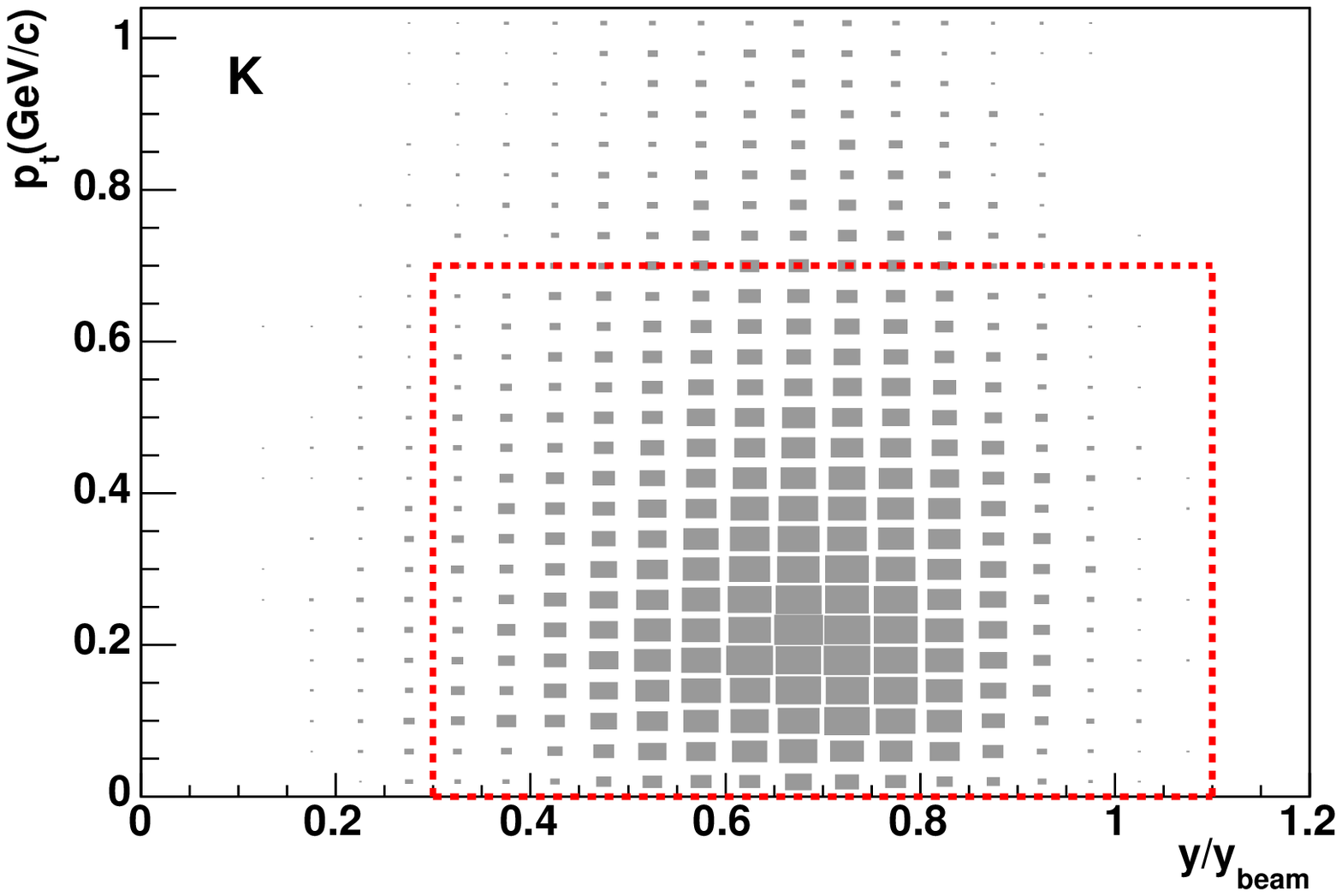}
\end{minipage}
\caption{\label{PS160} Phase space of mother particles. Grandmother energy range: \unit[80-400]{GeV}. Left panel: pions, right panel: kaons.
The dashed  box (red) indicates the most interesting phase space region which includes more than 90\% of this particles.}
\end{figure}

\begin{figure}[h]
\begin{minipage}[t]{0.49\textwidth}
\centering
\includegraphics[width =\textwidth, bb=  10 20 519 345,clip]{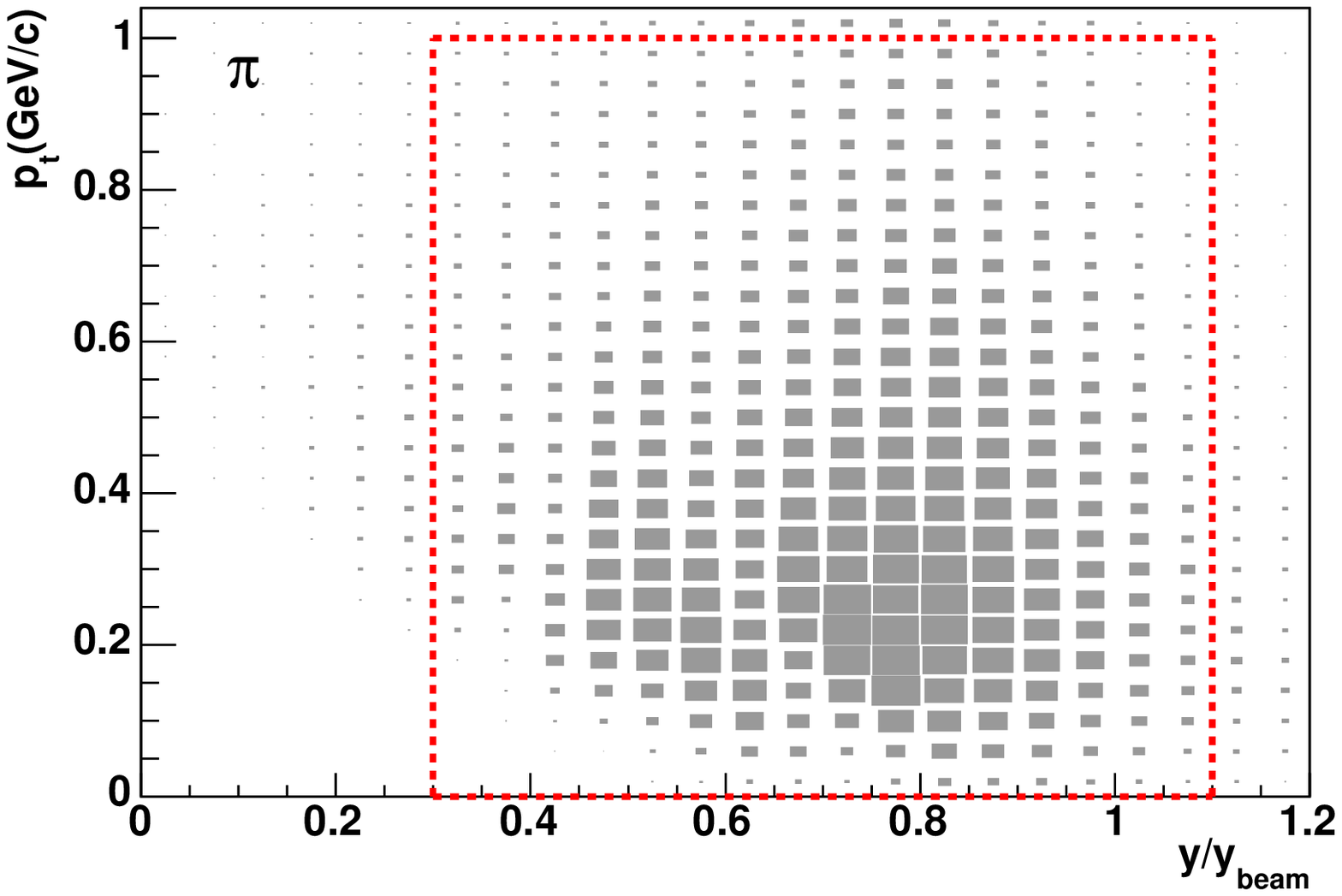}
\end{minipage}
\hfill
\begin{minipage}[t]{0.49\textwidth}
\centering
\includegraphics[width =\textwidth, bb=  10 20 519 345,clip]{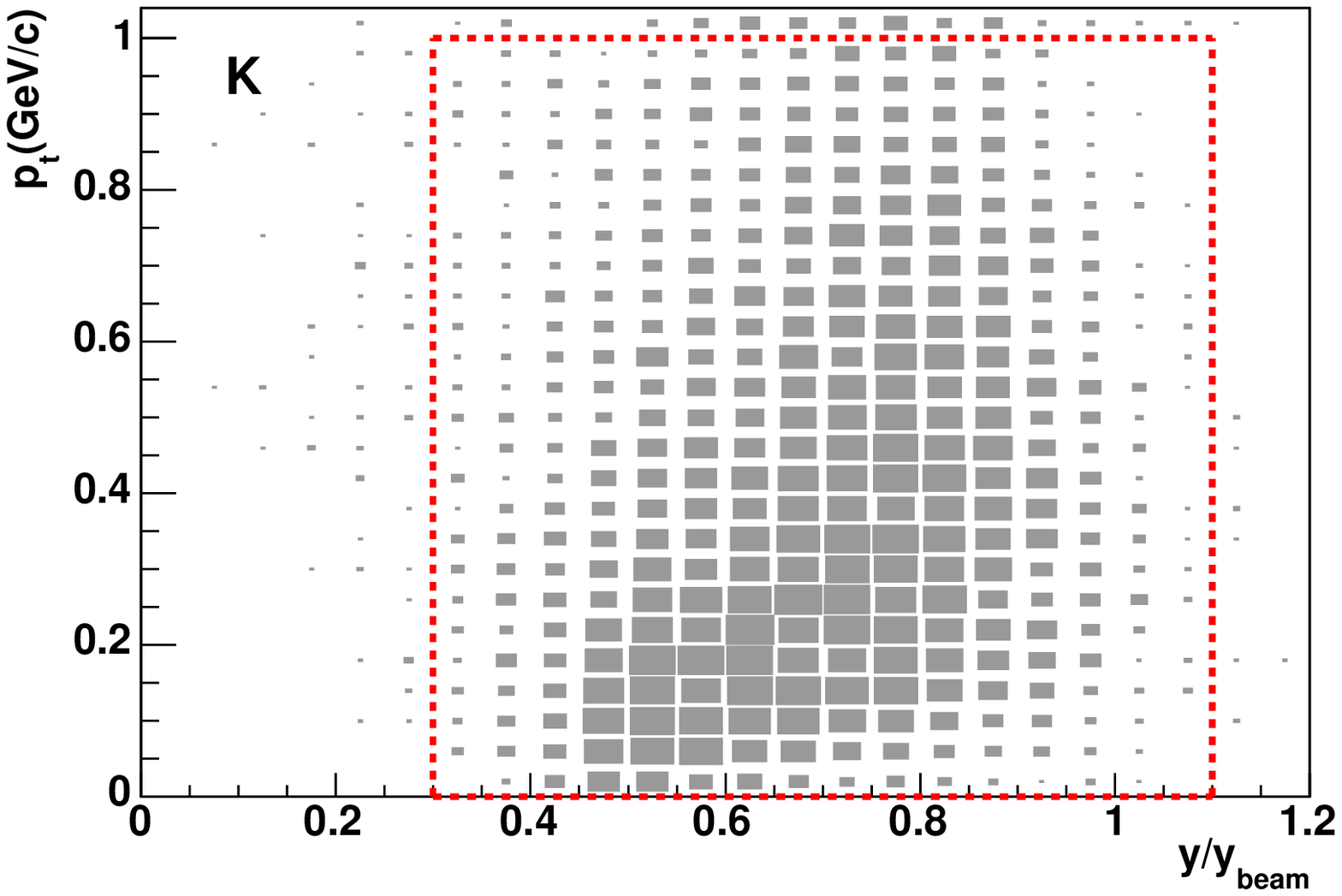}
\end{minipage}
\caption{\label{PS40} Same plots as in Fig.~\ref{PS160}, but the grandmother energy range is \unit[30-60]{GeV}.}
\end{figure}

The phase space regions of mother particles produced in the {\it last} interaction in EAS are shown in Fig.~\ref{PS160} and \ref{PS40}. We choose a nucleon as the grandmother particle with a mean energy of \unit[160]{GeV} (Fig.~\ref{PS160}) and \unit[40]{GeV} (Fig.~\ref{PS40}), respectively. The transverse momentum of the mother particle is plotted vs. its rapidity divided by the rapidity of the grandmother particle which is equivalent to the beam rapidity in fixed target reactions. On the left hand side this distribution is shown for pions as mother particles, on the right hand side for kaons.
The maximum of these distributions, which shows the most important phase space region for the muon production in EAS, is at $p_\perp\approx$~\unit[0.1]{GeV} and 0.7 in relative rapidity units (for a mean grandmother energy of \unit[160]{GeV}) and shifts to slightly higher $p_\perp$ but stays at the same rapidity for a mean grandmother energy of \unit[40]{GeV}. In both cases the distributions of pions and kaons are similar. For kaons higher particle transverse momenta are more important than for pions. 
The phase space regions of relevance to EAS are summarized in Tab.~\ref{ps_range} and indicated with the dashed (red) boxes in Fig.~\ref{PS160} and \ref{PS40}.

\begin{table}
\vspace*{-8mm}
\caption{\label{ps_range} Phase space regions of hadronic interactions 
relevant for muon production in EAS.}
\vspace*{2mm}
\centering
\begin{tabular}{|c|c|c|}
\hline average energy (GeV) & $y/y_{beam}$ & $p_\perp$ (GeV/c) \\
\hline
\hline 160                  & 0.3 - 1.1    & 0.0 - 0.7  \\
\hline  40                  & 0.3 - 1.1    & 0.0 - 1.0  \\
\hline
\end{tabular}
\vspace{0.5 cm}
\end{table}

\clearpage

\section{Existing fixed target data}

In contrast to collider experiments, in fixed target setups it is more straight forward to measure particles in forward direction. The energy range of most of the existing fixed target experiments has a broad overlap with the energy range of relevance to EAS. 

There exist a number of fixed target measurements employing a proton beam with low energy ($\leq$\unit[20]{GeV}) with a good phase space coverage, however, data at higher energy and for pion projectiles are very sparse.
 
\begin{figure}[h!]
\begin{minipage}[t]{0.5\textwidth}
\centering
\includegraphics[width =\textwidth, bb=30 30 532 508,clip]{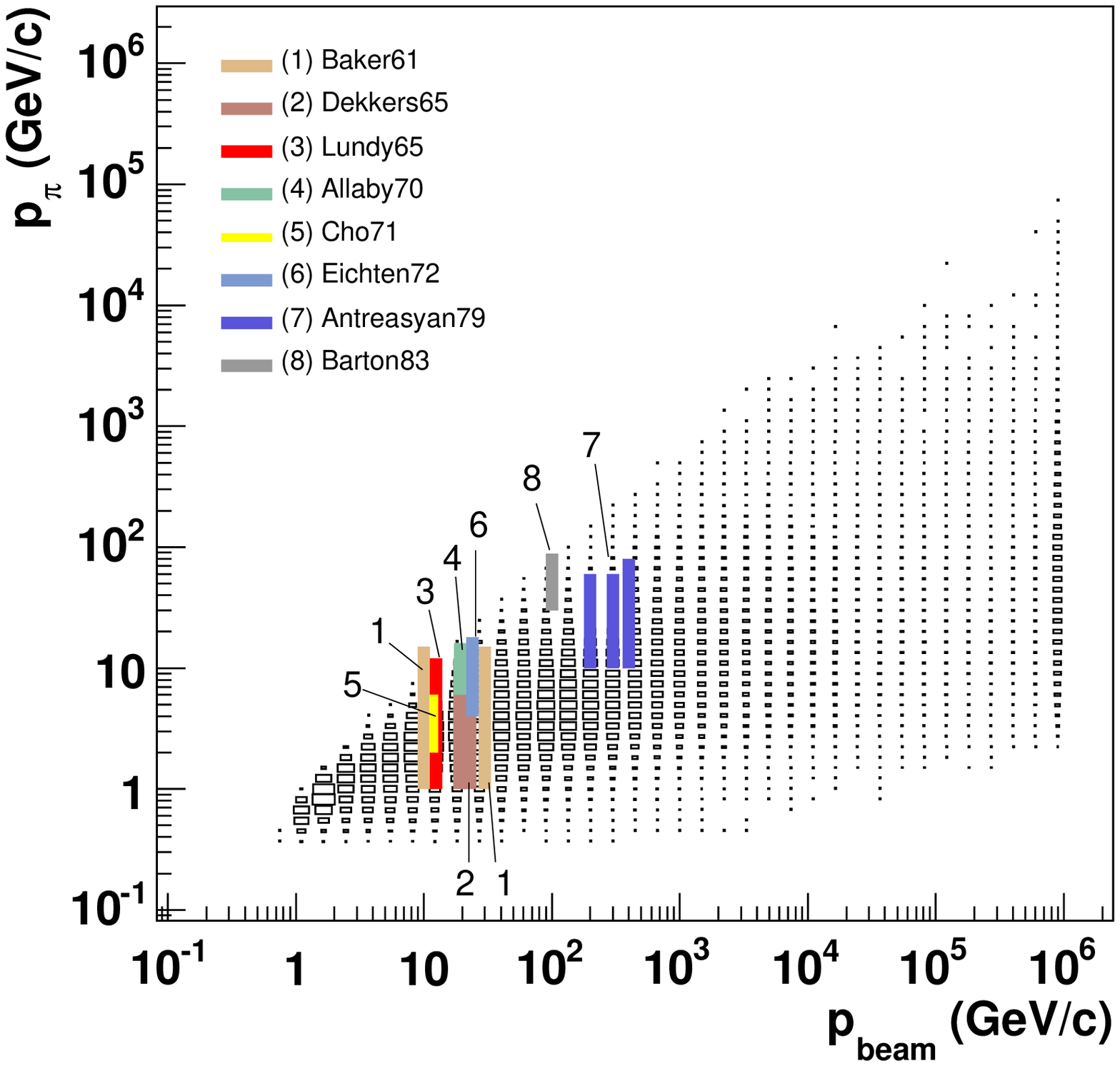} 
\end{minipage}
\hfill
\begin{minipage}[t]{0.5\textwidth}
\centering
\includegraphics[width =\textwidth, bb=30 30 532 508,clip]{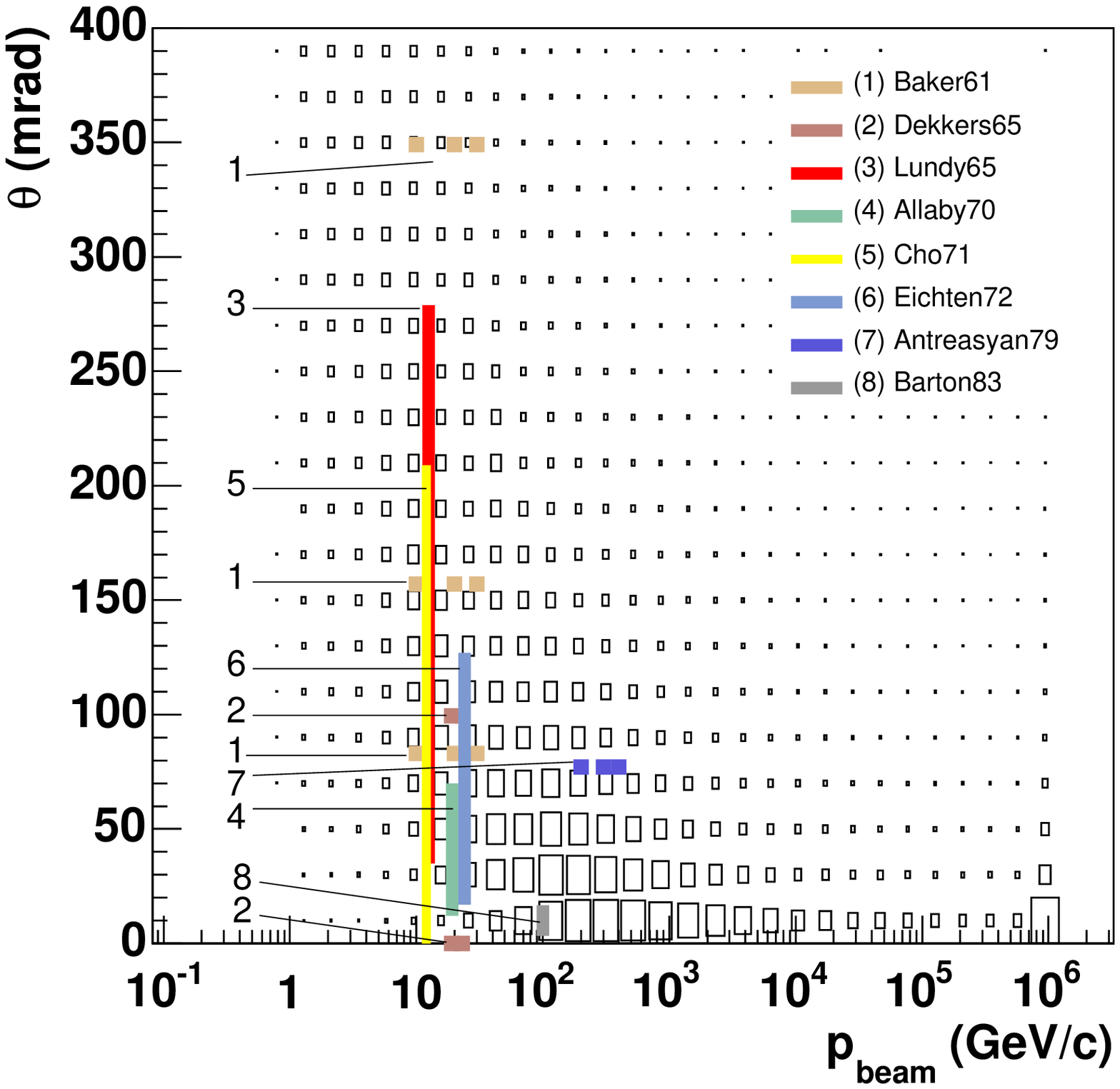}
\end{minipage}
\caption{\label{Dat_PS1} Coverage of the phase space regions of relevance to EAS (box histograms) by existing fixed target data using a proton beam and a beryllium or carbon target (shaded/colored regions).  Left panel: total momentum of secondary pions vs. total momentum of proton projectiles. Right panel: angle between beam and secondary particle momentum vs. beam momentum.}
\end{figure}

\begin{figure}[h!]
\begin{minipage}[t]{0.5\textwidth}
\centering
\includegraphics[width =\textwidth, bb=30 30 532 508,clip]{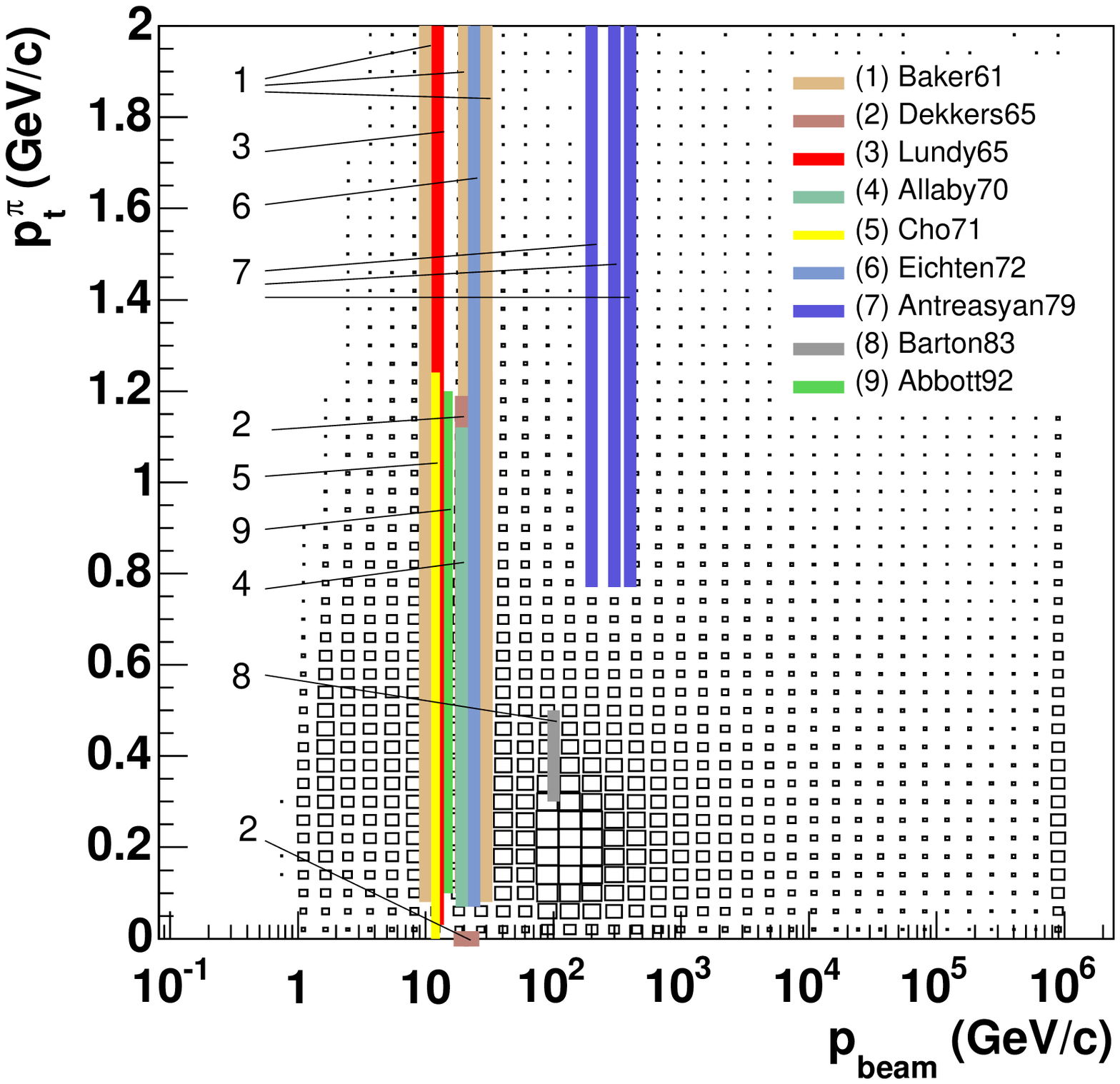}
\end{minipage}
\hfill
\begin{minipage}[t]{0.5\textwidth}
\centering
\includegraphics[width =\textwidth, bb=30 30 532 508,clip]{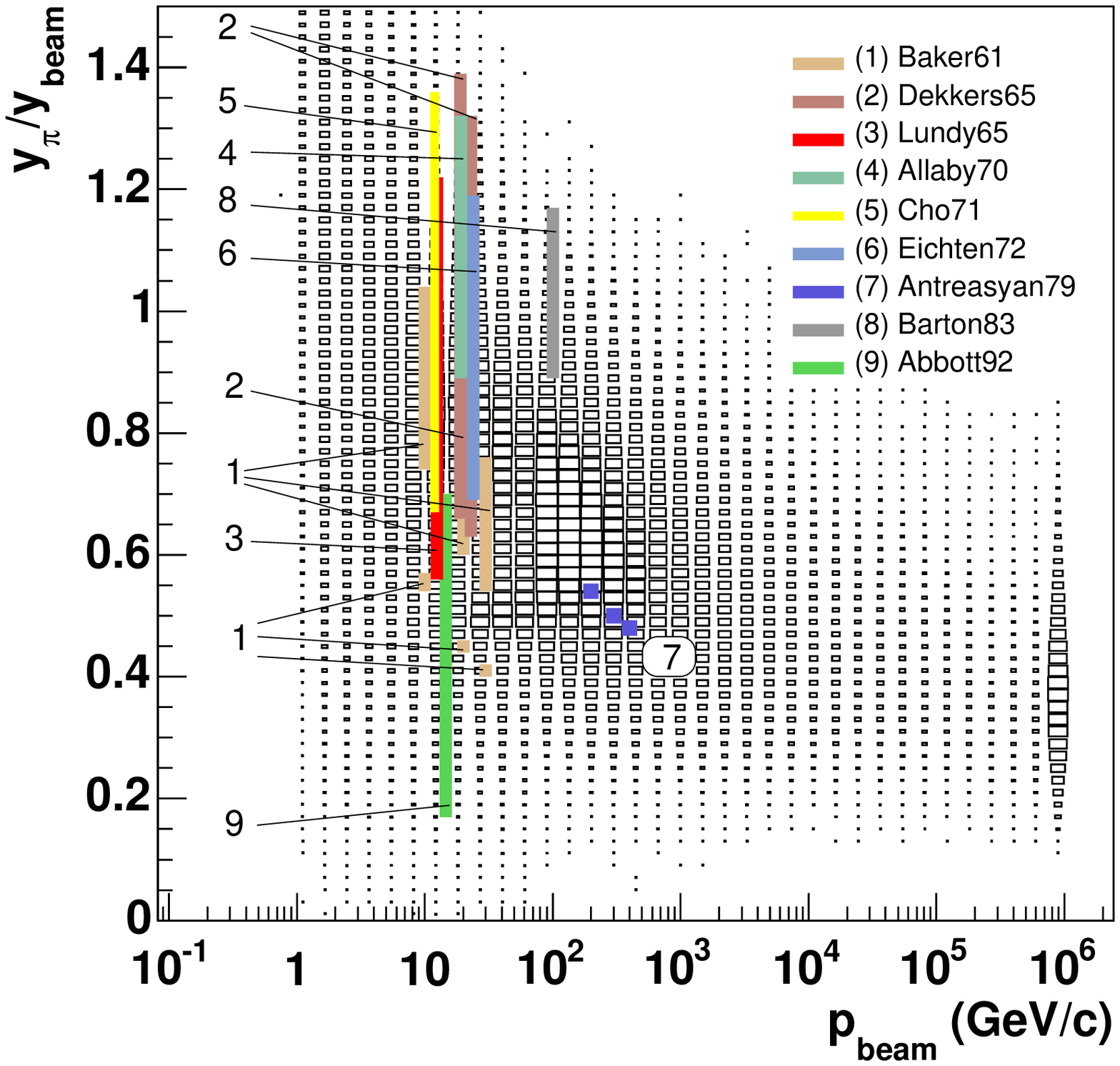}
\end{minipage}
\caption{\label{Dat_PS2} Compilation of the phase space regions covered by fixed target data given in transverse momentum and rapidity of secondary particles and the phase space regions covered by the $\theta-p_{\text{sec}}$ data (see Fig.~\ref{Dat_PS1}), whereas an approximate conversion of the covered phase space has been done.
Left panel: transverse momentum of secondary pions vs. total momentum of proton projectiles. Right panel: rapidity of secondary pions normalized by the beam rapidity vs. beam momentum.}
\end{figure}

In Fig.~\ref{Dat_PS1} a compilation of existing fixed target data is shown. Mostly the data on secondary particle distributions are given in the observables angle between beam and secondary particle momentum ($\theta$) and the momentum of the secondary particle itself ($p_{\text{sec}}$). Some measurements are also published using rapidity ($y$) and transverse momentum of secondary particles ($p_\perp$). For an easier comparison, in Fig.~\ref{Dat_PS2} all data sets are shown in the rapidity-transverse momentum plane, whereas the phase space regions covered by $\theta-p_{\text{sec}}$ measurements have been converted approximately.

In both figures the existing data are indicated by shaded (colored) regions, whereas the beam momentum is given as abscissa and the secondary particle observable as ordinate. The relevant phase space is shown as box histogram. In this case the grandmother momentum is used instead of the beam momentum for the abzissa. The size of the boxes indicates the relative importance of the beam and secondary particle momenta. 

The work of Eichten et al. \cite{Eichten72} has become a widely used reference data set. This experiment used a proton beam with a beam momentum of \unit[24]{GeV} and a beryllium target. The secondary particles (pions, kaons, protons) are measured in a broad angle range (\unit[17]{mrad} $<\theta<$ \unit[127]{mrad}) and in a momentum region from \unit[4]{GeV/c} up to \unit[18]{GeV/c}. Other measurements cover only a smaller part of the phase space of interest to EAS \cite{Baker61,Dekkers65,Lundy65,Allaby70,Cho71,Antreasyan79,Bartom83,Abbott92}.
A measurement for preparing the CERN neutrino beam experiments was performed by SPY/NA56 (NA52) \cite{Ambrosini98,Collazuol2000}. Therefore a \unit[450]{GeV} proton beam and a thick beryllium target was used. Because of the very limited angular range ($\theta \approx 0^{\circ}$) it is not included in Fig.~\ref{Dat_PS1} or Fig.~\ref{Dat_PS2}.

It is clear from this comparison (Figs.~\ref{Dat_PS1} and \ref{Dat_PS2}) that existing data do not cover the most important energy and secondary particle phase space regions.
However, there are three promising current ongoing data analyses. The HARP (PS214) experiment \cite{HARP} at the PS accelerator at CERN has taken p+C and $\pi$+C data at \unit[15]{GeV} in 2001. The MIPP (FNAL-E907) experiment \cite{MIPP} at Fermilab just started data taking in the beam energy range 5-\unit[120]{GeV} in 2005. Because of its excellent forward acceptance, it is also thought about using the NA49 detector \cite{NA49NIM} at the SPS accelerator at CERN in a follow-up experiment to perform similar measurements at higher energies (up to \unit[160]{GeV}) \cite{NA49future}.

\section{Conclusions and outlook}
Due to the interplay between decay and interaction of pions and kaons,
low energy hadronic interactions are very important for muon production
in extensive air showers. With increasing lateral distance the mean
energy of these interactions, which are mainly initiated by pions
and nucleons, decreases.    

The most important interaction energies and phase space regions are accessible for fixed target experiments. However, so far only measurements with protons as projectiles and with a very limited secondary particle phase space exist. The situation could be improved considerably, if data from fixed target experiments with large acceptance detectors such as HARP, NA49 and MIPP would be analysed for minimum bias collisions especially for $\pi$+C reactions.

Fixed target measurements can contribute to improving low energy interaction models. Due to the sensitivity of EAS to low energy hadron production, such improvements will increase the reliability of air shower simulations. 

Finally it should be noted that multi-detector installations such as KASCADE can be used to check the consistency of the simulation of hadronic interactions by comparing different observables of EAS and their correlation with model predictions.

\noindent 

{\bf Acknowledgements:} The authors thank Dieter Heck for 
many fruitful discussions and help with modifying CORSIKA to include 
the muon ancestor information.

\end {document}